\documentclass[pra,a4paper,twocolumn]{revtex4}

\usepackage{epsfig}
\usepackage{graphicx}
\usepackage{epstopdf}
\usepackage{amsmath, amsthm, amssymb}
\usepackage{comment}
\usepackage{subfigure}
\usepackage{algorithm}
\usepackage{algorithmic}
\usepackage{url}
\usepackage[toc,page]{appendix}
\usepackage{setspace}

\newcommand{\ket}[1]{| #1 \rangle}
\newcommand{\bra}[1]{\langle #1|}

\newcommand{\ip}[2]{\langle #1|#2 \rangle}

\newcommand{\tr}{\operatorname{tr}}

\newcommand{\be}{\begin{equation}}
\newcommand{\ee}{\end{equation}}
\newcommand{\bea}{\begin{eqnarray}}
\newcommand{\eea}{\end{eqnarray}}
\newcommand{\bes}{\begin{equation*}}
\newcommand{\ees}{\end{equation*}}
\newcommand{\beas}{\begin{eqnarray*}}
\newcommand{\eeas}{\end{eqnarray*}}

\newtheorem{thm}{Theorem}

\newtheorem{lem}[thm]{Lemma}

\begin{document}

\title{A lower bound on the probability of error in quantum state discrimination}

\date{\today}

\author{Ashley Montanaro}
\affiliation{Department of Computer Science, University of Bristol, Woodland Road, Bristol, BS8 1UB, U.K.}
\email{montanar@cs.bris.ac.uk}

\begin{abstract}
We give a lower bound on the probability of error in quantum state discrimination. The bound is a weighted sum of the pairwise fidelities of the states to be distinguished.
\end{abstract}


\maketitle


\section{Introduction}

The fact that non-orthogonal states are not perfectly distinguishable is a characteristic feature of quantum mechanics and the basis of the field of quantum cryptography. In this short note, we derive a quantitative lower bound on the indistinguishability of a set of quantum states.

The scenario we consider is that of quantum state discrimination: we are given a quantum system that was previously prepared in one of a known set of states, with known a priori probabilities, and must determine which state we were given with the minimum average probability of error. This fundamental problem was first studied by Helstrom~\cite{helstrom76} and Holevo~\cite{holevo73a} in the 1970s, and has since developed a vast literature (see \cite{chefles00} for a survey).

One can use efficient numerical techniques to determine this minimum average probability of error \cite{eldar03}, but a general closed-form expression appears elusive. We are therefore led to putting bounds on this probability. Such bounds have been useful in the study of quantum query complexity~\cite{childs07a} and in the security evaluation of quantum cryptographic schemes \cite{hayashi06}. However, prior to this work no lower bound based on the most natural ``local'' measure of distinguishability of the quantum states in question -- their pairwise fidelities -- was known.

The most general strategy for quantum state discrimination is given by a positive operator valued measure (POVM) \cite{peres95}, namely a set of positive operators $M = \{ \mu_i \}$ such that $\sum_i \mu_i = I$. The probability of receiving result $i$ from measurement $M$ on input of state $\rho$ is $\tr(\mu_i \rho)$. We define an ensemble $\mathcal{E}$ as a set of quantum states $\{\rho_i\}$, each with a priori probability $p_i$, and associate measurement outcome $i$ with the inference that we received state $\rho_i$. The average probability of error is then given by
\[ P_E(M,\mathcal{E}) = \sum_{i\neq j} p_j \tr(\mu_i \rho_j). \]
We mention some matrix-theoretic notation that we will require; for more details, see \cite{bhatia97}. For any matrix $M$ and real $p>0$, we define $\|M\|_p = \left(\sum_i \sigma_i(M)^p \right)^{1/p}$, where $\{\sigma_i(M)\}$ is the set of singular values of $M$. For $p \ge 1$ this is a matrix norm (known as the Schatten $p$-norm) and the case $p=1$ is known as the trace norm. As it only depends on the singular values of $M$, $\|M\|_p$ is invariant under pre- and post-multiplication by unitaries.

The fidelity (Bures-Uhlmann transition probability) between two mixed quantum states $\rho$, $\sigma$ can be defined in terms of the trace norm as $F(\rho,\sigma) = \|\sqrt{\rho} \sqrt{\sigma}\|_1^2$~\cite{uhlmann76,jozsa94}.

We can now state the main result of this paper as the following theorem.
\begin{thm}
\label{thm:main}
Let $\mathcal{E}$ be an ensemble of quantum states $\{\rho_i\}$ with a priori probabilities $\{p_i\}$. Then, for any measurement $M$,
\[ P_E(M,\mathcal{E}) \ge \sum_{i>j} p_i p_j F(\rho_i,\rho_j). \]
\end{thm}
We stress that this bound does not depend on the number of states in $\mathcal{E}$, nor their dimension.
Before proving this theorem, we compare the lower bound of this note with some related previous results.


\section{Previous work}

A classic result of Holevo and Helstrom \cite{helstrom76,holevo73a} gives the exact minimum probability of error that can be achieved when  discriminating between two states $\rho_0$ and $\rho_1$ with a priori probabilities $p$ and $1-p$:
\be \label{eqn:helstrom} \min_M P_E(M,\mathcal{E}) = \dfrac{1}{2}-\dfrac{1}{2}\|p \rho_0 - (1-p) \rho_1 \|_1. \ee
However, in the case where we must discriminate between more than two states, no such exact expression for the minimum $P_E(M,\mathcal{E})$ is known. Indeed, it appears that until recently the only known lower bound on $P_E(M,\mathcal{E})$ was a result of Hayashi, Kawachi and Kobayashi that gives a bound in terms of the individual operator norms of the states in $\mathcal{E}$ \cite{hayashi06}. A lower bound in terms of pairwise trace distances has very recently been given by Qiu \cite{qiu08}.

In the other direction, Barnum and Knill \cite{barnum02} developed a useful upper bound on the error probability, which is given by
\[ \min_M P_E(M,\mathcal{E}) \le 2 \sum_{i>j} \sqrt{p_i p_j} \sqrt{F(\rho_i,\rho_j)}. \]
It was pointed out by Harrow and Winter \cite{harrow06} that this leads to a worst-case upper bound on the number of copies required to achieve a specified probability of success of discriminating between a set of states whose pairwise fidelities are known to be bounded above by some constant. Similarly, Theorem \ref{thm:main} can be used to lower bound the number of copies required in an average-case setting. For example, assume that each pair of states $(\rho_i,\rho_j)$ has $F(\rho_i,\rho_j) \ge F$ for some $F$, that there are $n\ge 2$ equiprobable states to discriminate, and that we have $m$ copies of the state to test. Then
\[ P_E(M,\mathcal{E}) \ge \dfrac{1}{n^2} \sum_{i>j} F(\rho_i,\rho_j)^m \ge \frac{(n-1)F^m}{2n}, \]
so in order to achieve a error probability of at most $\epsilon$, we need to have access to at least
\[ m \ge \frac{\log_2(1/\epsilon)-2}{\log_2(1/F)}\]
copies of the test state.

Finally, we mention a related quantum state discrimination scenario that has been considered in the literature: \textit{unambiguous} state discrimination \cite{chefles00}. In this scenario, our measurement process is not allowed to make a mistake. That is, it is required that the measurement result is $i$ only if the input was state $i$. This can be achieved by allowing the possibility of failure, i.e.\ of outputting ``don't know''. Define $P^u_E(M,\mathcal{E})$ as the failure probability of an unambiguous measurement $M$ on ensemble $\mathcal{E}$. Zhang et al.\ have given a lower bound on this probability of failure in terms of the pairwise fidelity and $n$, the number of states to be discriminated \cite{zhang01}.
\[ P^u_E(M,\mathcal{E}) \ge \frac{2}{n-1} \sum_{i>j} \sqrt{p_i p_j} |\ip{\psi_i}{\psi_j}|. \]
Now let us turn to the proof of our main result.


\section{Proof of Theorem \ref{thm:main}}

We start by noting the following characterisation of a measurement based on that of Barnum and Knill~\cite{barnum02}. Decompose each state (weighted by its a priori probability) in terms of its eigenvectors as $p_i \rho_i = \sum_j \ket{e_{ij}}\bra{e_{ij}}$, where we fix the norm of each eigenvector $\ket{e_{ij}}$ as the square root of its corresponding eigenvalue $\lambda_{ij}$. Then define the matrix $S_i = \sum_j \ket{e_{ij}}\bra{j}$, and form the overall block matrix $S$ by writing the $S_i$ matrices in a row. That is, $S = \sum_{i,j} \ket{e_{ij}} \bra{i} \bra{j}$. If the states are not of equal rank, pad each matrix $S_i$ with zero columns so all the blocks are the same size.

Now perform the same task on an arbitrary measurement $M$. Perform the eigendecomposition of each measurement operator $\mu_i = \sum_j \ket{f_{ij}}\bra{f_{ij}}$ (again, the norm of each eigenvector is given by the square root of its corresponding eigenvalue), and form the matrix $N_i$ whose $j$'th column is $\ket{e_{ij}}$ (again, padding with zero columns if necessary). Write these matrices in a row to give $N = \sum_{i,j} \ket{f_{ij}} \bra{i} \bra{j}$. As $\sum_i \mu_i = I$, it is immediate that $N N^\dag = I$.

Set $A = N^\dag S$. $A$ is made up of blocks $A_{ij} = N_i^\dag S_j$. It is easy to verify that the probability of error of the measurement is completely determined by $A$:
\[ \|A_{ij}\|_2^2 = \tr((N_i N_i^\dag)(S_j S_j^\dag)) = p_j \tr(\mu_i \rho_j), \]
so the squared 2-norm $\|A_{ij}\|_2^2$ gives the probability of receiving state $j$ and identifying it as state $i$, and we have $P_E(M,\mathcal{E})=\sum_{i\neq j} \|A_{ij}\|_2^2$.

Our proof rests on the fact that on the one hand $A^\dag A = S^\dag N N^\dag S = S^\dag S$, and on the other the pairwise fidelities of the states in $\mathcal{E}$ can also be obtained from $S^\dag S$. Indeed, consider the $(i,j)$'th block of this matrix, $(S^\dag S)_{ij} = S_i^\dag S_j$. If the states in $\mathcal{E}$ are all pure (say $\rho_i = \ket{\psi_i}\bra{\psi_i}$), then each block is a $1 \times 1$ matrix $(S^\dag S)_{ij} = \sqrt{p_i} \sqrt{p_j} \ip{\psi_i}{\psi_j}$. That is, $S^\dag S$ is the Gram matrix of the states in $\mathcal{E}$ \cite{bhatia97}, scaled by their a priori probabilities.

More generally, we have $S_i S_i^\dag = p_i \rho_i$. This implies that, by the polar decomposition of $S_i$, $S_i = \sqrt{p_i \rho_i}\,U$ for some unitary $U$. Thus, for some unitary $U$ and $V$,
\begin{align*}
\|S_i^\dag S_j\|_1^2 &= \| U^\dag \sqrt{p_i \rho_i} \sqrt{p_j \rho_j}\,V \|_1^2 = p_i p_j \| \sqrt{\rho_i} \sqrt{\rho_j} \|_1^2\\
&= p_i p_j F(\rho_i,\rho_j),
\end{align*}
where the second equality follows from the unitary invariance of the trace norm.

Our approach, following \cite{barnum02}, will be to use these facts to lower bound the sum $\sum_{j \neq i} \|A_{ij}\|_2^2$ for a fixed $i$ in terms of the entries of $A^\dag A$, and then to sum over $i$. We will require two matrix norm inequalities. The first appears to be new, and the second was proven by Bhatia and Kittaneh using a duality argument \cite{bhatia90}; we give a simple direct proof for completeness.
\begin{lem}
\label{lem:normbound}
Let $A$, $B$, $C$, $D$ be square matrices of the same dimension. Then
\[ \|AB + CD\|_1^2 \le (\|A\|_2^2+\|D\|_2^2)(\|B\|_2^2+\|C\|_2^2). \]
\end{lem}
\begin{proof}
Perform the polar decomposition $CD=PU$ for some positive semidefinite $P$ and unitary $U$. Then
\begin{align*}
\|AB+CD\|_1 &= \|AB+PU\|_1 = \|AB+P^\dag U\|_1\\
&= \|ABU^\dag+P^\dag\|_1 = \|ABU^\dag + UD^\dag C^\dag\|_1,
\end{align*}
where the third equality follows from the unitary invariance of the trace norm. Writing this as the product of two block matrices,
\begin{align*}
\|&AB+CD\|_1^2\\
&= \|\begin{pmatrix} A & U\!D^\dag \end{pmatrix} \begin{pmatrix} BU^\dag & C^\dag \end{pmatrix}^\mathrm{T}\|_1^2 \\
&\le \|A A^\dag + U D^\dag D U^\dag\|_1 \|U B^\dag B U^\dag + C C^\dag\|_1\\
&\le (\|A A^\dag\|_1 + \|U D^\dag D U^\dag\|_1)(\|U B^\dag B U^\dag\|_1 + \|C C^\dag\|_1)\\
&= (\|A\|_2^2+\|D\|_2^2)(\|B\|_2^2+\|C\|_2^2),
\end{align*}
where the first inequality is the Cauchy-Schwarz inequality for unitarily invariant norms \cite{bhatia97} and the second is the triangle inequality.
\end{proof}
\begin{lem}[Bhatia and Kittaneh \cite{bhatia90}]
\label{lem:partitioned}
Let $M$ be a block matrix $M=(M_1 \dots M_n)$. Then $\|M\|_1^2 \ge \sum_i \|M_i\|_1^2$.
\end{lem}

\begin{proof}
Let $N_i$ be the matrix given by replacing all blocks in $M$ other than block $i$ with zeroes. Then it is easy to see that
\[ M^\dag M = \sum_i N_i^\dag N_i \]
and also that $\|M\|_1 = \|\sqrt{M^\dag M}\|_1$, $\|M_i\|_1 = \|\sqrt{N_i^\dag N_i}\|_1$. Thus
\begin{align*}
\|M\|_1^2 &= \|\sqrt{\sum_i N_i^\dag N_i}\|_1^2 = \|\sum_i N_i^\dag N_i\|_{1/2}\\
& \ge \sum_i \|N_i^\dag N_i\|_{1/2} = \sum_i \|\sqrt{N_i^\dag N_i}\|_1^2 = \sum_i \|M_i\|_1^2,
\end{align*}
where the inequality in the second line can be proven easily by a majorisation argument \cite{bhatia97}, and is given explicitly as Lemma 1 of \cite{bhatia00}.
\end{proof}
We now return to the proof of Theorem \ref{thm:main}. Group the blocks of $A$ into four ``super-blocks'' as follows:
\[ A = \begin{pmatrix}
\begin{pmatrix} A_{11} \end{pmatrix} & \begin{pmatrix} A_{12} & \dots & A_{1n} \end{pmatrix}\\
\begin{pmatrix} A_{21}\\ \vdots \\ A_{n2} \end{pmatrix} &
\begin{pmatrix} A_{22} & \dots & A_{2n}\\ \vdots & \ddots & \vdots \\ A_{n2} & \dots & A_{nn} \end{pmatrix}
\end{pmatrix}. \]
Now define a new $2 \times 2$ block matrix $B$ by setting block $B_{ij}$ to be the corresponding super-block in the above decomposition of $A$, appending rows and/or columns of zeroes to each of these blocks such that each block in $B$ is square. Super-block $A_{12}$ is thus the first row of block $B_{12}$. Consider the product $B^\dag B$ with the same block structure. One can verify that the first row of the block $(B^\dag B)_{12}$ is equal to the submatrix of $A^\dag A$ defined as $T = ((A^\dag A)_{12} \dots (A^\dag A)_{1n})$, and the remaining rows in this block are zero. We therefore have $\|(B^\dag B)_{12}\|_1 = \|T\|_1$. Using Lemma \ref{lem:partitioned} followed by Lemma \ref{lem:normbound} gives
\begin{align*}
\sum_{i>1} \|(A^\dag A)_{1i}\|_1^2 &\le \|T\|_1^2 = \|B_{11}^\dag B_{12} + B_{21}^\dag B_{22}\|_1^2\\
&\le (\|B_{11}\|_2^2+\|B_{22}\|_2^2)(\|B_{12}\|_2^2+\|B_{21}\|_2^2)\\
&\le \|B_{12}\|_2^2+\|B_{21}\|_2^2\\
&= \sum_{i>1} \|A_{1i}\|_2^2 + \|A_{i1}\|_2^2,
\end{align*}
where we use the fact that $\sum_{i,j} \|B_{ij}\|_2^2 = \sum_{i,j} \|A_{ij}\|_2^2=1$ in the final inequality. We may now proceed to obtain corresponding inequalities for the other rows of $A$ by permuting its rows and columns. Summing these inequalities, and noting that each off-diagonal element of $A$ appears twice in total, gives
\begin{align*}
P_E(M, \mathcal{E}) &= \sum_{i \neq j} \|A_{ij}\|_2^2 \ge \sum_{i>j} \|(A^\dag A)_{ij}\|_1^2\\
&= \sum_{i>j} \|(S^\dag S)_{ij}\|_1^2 = \sum_{i>j} p_i p_j F(\rho_i,\rho_j)
\end{align*}
and the proof is complete.


\section{Concluding remarks}

We have given a lower bound on the probability of error in quantum state discrimination that depends only on the pairwise fidelities of the states in question and is appealingly similar to a known upper bound of Barnum and Knill \cite{barnum02}. We close by commenting on the tightness of this bound.

It can be seen by comparing Theorem \ref{thm:main} with the Helstrom bound (\ref{eqn:helstrom}) that the lower bound of this paper is not always tight, even for two states, but is nevertheless close to optimal (in some sense). Consider a pair of identical states $\rho_0 = \rho_1 = \rho$ for some arbitrary $\rho$. Then, by (\ref{eqn:helstrom}),
\[ \min_M P_E(M, \mathcal{E}) = \frac{1}{2} - \frac{1}{2} \| (p - (1-p)) \rho \|_1 = \frac{1}{2} - |p-\frac{1}{2}|, \]
whereas Theorem \ref{thm:main} guarantees only a weaker lower bound of
\[ \min_M P_E(M, \mathcal{E}) \ge p(1-p). \]
On the other hand, this lower bound cannot be improved by any constant factor $\alpha > 1$ without violating (\ref{eqn:helstrom}).


{\bf Note added.} Following the completion of this work, I became aware of recent work by Qiu \cite{qiu08}, which obtains a lower bound on $P_E(M, \mathcal{E})$ in terms of pairwise trace distances. For an ensemble of 2 states, this bound reduces to the Holevo-Helstrom quantity (\ref{eqn:helstrom}).


\section*{Acknowledgements}

This work was supported by the EC-FP6-STREP network QICS. I would like to thank Richard Jozsa for helpful discussions on the subject of this paper, and Daowen Qiu for bringing reference \cite{qiu08} to my attention.


\end{document}